\documentclass[preprint,showpacs,preprintnumbers,amsmath,amssymb,natbib]{revtex4}

\usepackage{graphicx}% Include figure files
\usepackage{epsfig}		
\usepackage{dcolumn}% Align table columns on decimal point
\usepackage{bm}% bold math
\def\0{\mbox{\tiny $0$}}
\def\1{\mbox{\tiny $1$}}
\def\2{\mbox{\tiny $2$}}
\def\3{\mbox{\tiny $3$}}
\def\4{\mbox{\tiny $4$}}
\def\5{\mbox{\tiny $5$}}
\def\6{\mbox{\tiny $6$}}
\def\7{\mbox{\tiny $7$}}
\def\8{\mbox{\tiny $8$}}
\def\9{\mbox{\tiny $9$}}

\def\f14{\mbox{\tiny $\frac{1}{4}$}}

\def\bb#1{\mbox{\footnotesize $(#1)$}}
\begin{document}
%
%%%%%%%%%%%%%%%%%%%%%%%%%%%%%%%% PAPER %%%%%%%%%%%%%%%%%%%%%%%%%%%%%%%%%%%%%

\title{Cosmological neutrino entropy changes due to flavor statistical mixing}
%\title{Corrections to cosmological neutrino temperatures due to the flavor associated Von Neumann entropy}
%\title{Flavor weighted energy implications to the cosmological neutrino mass}
\author{A. E. Bernardini}
\email{alexeb@ufscar.br}
\altaffiliation[On leave of absence from]{~Departamento de F\'{\i}sica, Universidade Federal de S\~ao Carlos, PO Box 676, 13565-905, S\~ao Carlos, SP, Brasil.}
\affiliation{Departamento de F\'isica e Astronomia, Faculdade de Ci\^{e}ncias da
Universidade do Porto, Rua do Campo Alegre 687, 4169-007, Porto, Portugal.}%\author{M. M. Guzzo}

\date{\today}

\renewcommand{\baselinestretch}{1.3}

\begin{abstract}
Entropy changes due to delocalization and decoherence effects should modify the predictions for the cosmological neutrino background (C$\nu$B) temperature when one treats neutrino flavors in the framework of composite quantum systems.
Assuming that the final stage of neutrino interactions with the $\gamma e^{-}e^{+}$ radiation plasma before decoupling works as a measurement scheme that projects neutrinos into flavor quantum states, the resulting free-streaming neutrinos can be described as a statistical ensemble of flavor-mixed neutrinos.
Even not corresponding to an electronic-flavor pure state, after decoupling the statistical ensemble is described by a density matrix that evolves in time with the full Hamiltonian accounting for flavor mixing, momentum delocalization and, in case of an open quantum system approach, decoherence effects.
Since the statistical weights, $w$, shall follow the electron elastic scattering cross section rapport given by $0.16\,w_{e} = w_{\mu} = w_{\tau}$, the von-Neumann entropy will deserve some special attention.
Depending on the quantum measurement scheme used for quantifying the entropy, mixing associated to dissipative effects can lead to an increasing of the flavor associated von-Neumann entropy for free-streaming neutrinos.
The production of von-Neumann entropy mitigates the constraints on the predictions for energy densities and temperatures of a cosmologically evolving isentropic fluid, in this case, the cosmological neutrino background.
Our results states that the quantum mixing associated to decoherence effects are fundamental for producing an additive quantum entropy contribution to the cosmological neutrino thermal history.
According to our framework, it does not modify the predictions for the number of neutrino species, $N_{\nu} \approx 3$.
It can only relieve the constraints between $N_{\nu}$ and the neutrino to radiation temperature ratio, $T_{\nu}/T_{\gamma}$, by introducing a novel ingredient to re-direct the interpretation of some recent tantalizing evidence than $N_{\nu}$ is significantly larger than by more than $3$.
\end{abstract}

\pacs{03.65.Ta, 14.60.Pq, 98.80.-k}

\keywords{flavor mixing and oscillation, neutrino, cosmology, density matrix}

\maketitle
\newpage
\renewcommand{\baselinestretch}{1.4}

The question of late-time entropy production that leads to changes on the cosmological neutrino background (C$\nu$B) temperature \cite{01,02,03,04} has been recently posed as theoretical puzzle on the foundations of the cosmological standard model.
The textbook literature sets that, due to the electron-positron ($e^+e^-$) annihilations which heated the background radiation after neutrino-radiation decoupling, the C$\nu$B temperature should be lower by a factor $T_{\nu\0}/T_{\gamma\0} = (4/11)^{1/3}$ when compared to the cosmological microwave background (CMB) temperature, where $T_{\nu\0}$ and $T_{\gamma\0}$ are, respectively, the C$\nu$B and CMB temperatures at present.

Speculative factors which may cause some slight departure from the standard value of $T_{\nu\0}/T_{\gamma\0} = (4/11)^{1/3}$ produce an overall increasing of the neutrino energy density of an order of $1\%$, i. e. a tiny effect when included in practical calculations.
It is originated either from corrections due to finite-temperature quantum field theories, which lead to an additional slight heating of the neutrinos \cite{144}, or from a secondary heating due to $e^+e^-$ annihilation prior to decoupling \cite{02}.

The questions posed in this letter are therefore how flavor mixing effects could introduce additional corrections to $T_{\nu\0}/T_{\gamma\0}$, which become effective some time after decoupling, deep inside the free-streaming regime, and how it affects the constraints over the number of neutrino species.
Flavor related von-Neumann entropies are therefore relevant in the context of describing neutrino flavor oscillations in the framework of composite quantum system \cite{Bernardini}.
Any relative  contribution to the entropy changes of a cosmologically evolving isentropic fluid, namely the C$\nu$B, could modify the predictions for its corresponding temperature which would affect the rapport to the energy densities.
Since the von-Neumann entropy for a composite quantum system of well-defined flavor quantum numbers is obtained in the framework of the generalized theory of quantum measurement, we extend such a discussion to the arena of cosmological neutrinos and we provide a general and functional characterization of the corresponding flavor associated entropies.

Let us recapitulate that the main difference between $T_{\nu}$ and $T_{\gamma}$ must arise from $e^+e^-$ annihilations after neutrino decoupling from the $\gamma e^+e^-$ radiation plasma.
Prior to annihilations, the total entropy density that includes all the ultra-relativistic fermionic and bosonic species is given by \cite{Dodelson}
\begin{equation}
s_i (a) = \frac{2\pi^2}{45} \, T_r^{3} \left[ g_{\gamma} + \frac{7}{8} \left(g_{e} + g_{\nu}\right)\right],
\label{alex1}
\end{equation}
where $g_{\gamma} = 2$, $g_{e} = 4$, and $g_{\nu} = 6$ are the respective numbers of degrees of freedom for photon, electron/positron and neutrino/antineutrino according to the Standard Model, and $T_r$ is the temperature of the radiation plasma.
After annihilations, the electrons and positrons have gone away and the photon and neutrino temperatures are no longer
identical.
The total entropy density is thus given in terms of the above defined temperatures, $T_{\nu}$ and $T_{\gamma}$, by
\begin{equation}
s_f (a) = \frac{2\pi^2}{45} \,  T_{\nu}^{3} \left[ g_{\gamma}\, \frac{ T_{\gamma}^{3}}{ T_{\nu}^{3}} + \frac{7}{8} g_{\nu}\right]
\label{alex2}.
\end{equation}
Assuming the isentropic evolution over Eqs.~(\ref{alex1}) and (\ref{alex2}), one sets $a^{3} \, s_i(a) = a^{3} \, s_f(a)$ and easily obtains the temperature ratio $T_{\nu\0}/T_{\gamma\0} = (4/11)^{1/3}$, which can be substituted into the energy and particle number density definitions, $\rho$ and $n$, in order to give $\rho_{\nu}/\rho_{\gamma} = N_{\nu} (7/8) (4/11)^{4/3}$ and $n_{\nu}/n_{\gamma} = 3/11$, where $N_{\nu}$ is the number of neutrino species \cite{Dodelson}.
One should notice that the value of $g_{\nu} = 6$ is introduced in the absence of right-handed neutrinos.
%If neutrinos possess a Majorana mass, then right-handed neutrinos do not necessarily exist, but if they have a Dirac mass, both left- and right-handed particles could be included into the above calculations (by setting $g = 12$).
As the right handed neutrinos are not produced in the early universe (the masses are so small as to be irrelevant) so only one spin state is produced and $g = 6$ \footnote{Indeed $g = 12$ is already completely ruled out by cosmological data.}.
Meanwhile, it is straightforward to verify that the number of $\nu$ degrees of freedom do not affect the rapport between $T_{\nu}$ and $T_{\gamma}$.

Moreover, as usually noticed in the literature, the temperature ratio of radiation and neutrinos is assumed to be constant.
Once neutrinos are massive they may enter the non-relativistic regime below a scale factor $a$ of about $10^{-4}$.
The reference \cite{Ma94} provides us with a suitable explanation for the consequence of finite neutrino thermal speed: the initial phase-space density for massive neutrinos is a relativistic Fermi-Dirac distribution, preserved from the time when the neutrinos decoupled in the early universe.
Decreasing the temperature with time is compensated by relating proper momentum to comoving momentum.
Therefore, ignoring perturbations, the present-day distribution for massive neutrinos is the relativistic Fermi-Dirac - not the equilibrium nonrelativistic distribution - because the phase space-distribution was preserved after neutrino decoupling.
The neutrinos distribution function is similar to that for a massless particle and
neutrinos temperature scales as $a^{-1}$.

Assuming that free-streaming flavor oscillating neutrinos evolve isentropically straightforward to the era of low temperatures, the maximal free streaming neutrino entropy may change due to decoherence effects.
It is parameterized by $\delta S_{\mbox{\tiny VN}}$ into the following relation for neutrino temperatures,
\begin{equation}
g_{\nu} \left(\frac{7}{8}\right) \frac{2\pi^2}{45}\, T^{3}_{\nu} = g^{\prime}_{\nu} \left[\left(\frac{7}{8}\right)  \frac{2\pi^2}{45}\, T^{\prime \,3}_{\nu} \pm n_{\nu}\, \delta S_{\mbox{\tiny VN}}\right].
\label{xxxx}
\end{equation}
By observing that $n_{\nu} = (9/5)\,\pi^{-2}\, T^{\prime 3}_{\nu}$, and assuming that $g^{\prime}_{\nu} = g_{\nu}$, i. e. that the number of degrees of freedom is conserved, Eq.~(\ref{xxxx}) results in
\begin{equation}
T_{\nu} = T^{\prime}_{\nu} \left(1 \pm \frac{324}{7 \pi^{4}}\,\delta S_{\mbox{\tiny VN}} \right)^{\frac{1}{3}}
\label{xxx}
\end{equation}
where, as we shall notice, $\delta S_{\mbox{\tiny VN}}$ stands for the associated von-Neumann entropy per flavor quantum ensemble in a volume $dp^{3}\,dq^{3}$ of the phase-space, so that $n_{\nu} \,\delta S_{\mbox{\tiny VN}}$ corresponds to the total von-Neumann entropy density.

The modifications introduced by Eq.~(\ref{xxx}) are guaranteed by the subadditivity of entropies associated to density matrices that describe independent quantum systems $A$ and $B$, i. e. systems with completely or partially uncorrelated quantum properties.
For cosmological neutrino ensembles, the quantum property related to the system $A$ is approximated by the momentum, $p$, which is read as a quantum phase-space element through its relation to the classical momentum established by the distribution function, $f(p)$, utilized for computing averaged quantities like $\rho_{\nu}$ and $n_{\nu}$ \cite{Dodelson}.
The quantum property related to the system $B$ is obviously the flavor quantum number.
The triangle inequality given by
\begin{equation}
|S_A -S_B|\,\leq \, S_{AB} \, \leq \, |S_A + S_B|,
\label{sdsd}
\end{equation}
sets upper and lower limits for the total entropy, $S_{AB}$, that describes a composite quantum system with the abovementioned characteristics.
While according to the Shannon's theory the entropy of a composite system can never be lower than the entropy of any of its parts \cite{Breuer}, in quantum mechanics the triangle inequality sets that the entropy of the joint system can be less than the sum of the entropy of its components due to the possibility of entanglement.
It happens when the localization character introduced by the momentum distribution function, $f(p)$, changes the pattern of flavor oscillations, as an intrinsic decoherence mechanism promoted by some dynamics of the delocalization effect similar to  those caused by the space-time evolution of mass-eigenstate wave-packets.
In parallel, the right-hand inequality can be interpreted as saying that the entropy of a composite system is maximized when its components are completely uncorrelated, i. e. when the entanglement disappears.

Turning back to the main point of our analysis, let us report about some foundations on quantum statistics in order to quantify $\delta S_{\mbox{\tiny VN}}$ into Eq.~(\ref{xxx}).
The density matrix representation of a composite quantum system of three flavor species, namely $e$, $\mu$ and $\tau$, is given by
\begin{equation}
\rho\bb{t} \equiv \rho = \sum_{\alpha = e,\mu,\tau}w_{\alpha} \, M^{\alpha}_{(t)} ~~\mbox{with} ~~ \sum_{\alpha = e,\mu,\tau} w_{\alpha} = 1,
\label{eq05}
\end{equation}
where $w_{\alpha}$ are the statistical weights, and $M^{\alpha}_{(t)}$ are the $\alpha$-flavor projection operators which depend on the flavor associated mixing angles and are constrained by the unitarity condition $\sum_{\alpha} M^{\alpha}_{(t)} = \mathbf{1}$.

The von-Neumann entropy provides an important functional defined in terms of the density matrix as
\begin{equation}
S\bb{\rho} = - \mbox{Tr}\{\rho\, \ln\bb{\rho}\},
\label{eq26Z}
\end{equation}
where the Boltzmann constant, $k_{B}$, was set equal to unity.
The entropy $S\bb{\rho}$ quantifies the departure of a composite quantum system from a pure state, i. e. it implicitly measures the entanglement of an ensemble of flavor states describing a given finite system.
As one can expect, quantum measurements induce modifications on the the von-Neumann entropy of the system.
The entropy change due to a {\em non-selective} measurement scheme described by {\em operations} parameterized by the projection operators $M^{\alpha}_{(0)}$ \cite{Breuer} is given by
\begin{equation}
\Delta S = S\bb{\rho^{\prime}} - S\bb{\rho} \geq 0,
\label{eq26A}
\end{equation}
where
\begin{equation}
S\bb{\rho^{\prime}} =
S\left(\sum_{\alpha} {P^{\alpha}_{(t)} \rho_{\alpha}}\right),
\label{eq26B}
\end{equation}
with
%\begin{equation}
$\rho_{\alpha} = \left(P^{\alpha}_{(t)}\right)^{-1} \, M^{\alpha}_{(0)}\, \rho \, M^{\alpha}_{(0)},$
%\label{eq14}
%\end{equation}
and where $P^{\alpha}_{(t)}$ are the probabilities of measuring $\alpha$-flavor eigenstates at time $t$.
In terms of the density matrix, one has
\begin{eqnarray}
P^{\alpha}_{(t)}   = \mbox{Tr}\{M^{\alpha}_{(0)}\, \rho\} &=& \sum_{\beta = e,\mu,\tau}{w_{\beta} \, \mbox{Tr}\{M^{\alpha}_{(0)}\,M^{\beta}_{(t)}\}} = %\nonumber\\&=&
\sum_{\beta = e,\mu,\tau}{w_{\beta} \,\mathcal{P}_{\alpha\rightarrow \beta}\bb{t}},
\label{eq06}
\end{eqnarray}
with $\mathcal{P}_{\alpha\rightarrow \beta}\bb{t} =  |\langle \nu^{\beta}_{(0)}\mid\nu^{\alpha}_{(t)} \rangle|^{\2}$ describing the $\alpha$- to $\beta$-flavor conversion probabilities in the single-particle quantum mechanics framework.

We assume that $\delta S_{\mbox{\tiny VN}}$ quantifies the level of flavor-mixing during the evolution of cosmological neutrino ensembles either from pure states to statistical mixtures, or from statistical mixtures to maximal statistical mixtures.
It results into a kind of late-time entropy production.

The above conceptions related to the von-Neumann entropy bring up important insights into the scope of distinguishing measurement procedures \cite{Breuer} and quantifying the degree of mixture of statistical ensembles.
In order to distinguish possible decoherence effects, i. e. those caused by dissipative mechanisms (extrinsic decoherence) from those due to delocalization characteristics (intrinsic decoherence), we shall parameterize $\delta S_{\mbox{\tiny VN}}$ by two different ways.
Both effects lead to increasing entropy density values after neutrino decoupling.

The cosmological standard model prescription for neutrino decoupling in the early universe sets that the three neutrino species ($e$, $\mu$, $\tau$) are kept in thermal contact with the radiation plasma through the elastic scattering process with background electrons(positrons).
Different flavor neutrinos ($e$, $\mu$, $\tau$) coexist with the {\em same averaged temperature}: neutrinos corresponding to the same volume of the phase space (that is constrained by some momentum distribution), reach the thermal equilibrium through {\em measurement} schemes produced by the elastic scattering.
The proportion between the corresponding cross sections, $\sigma_{\nu}$, is given by
\begin{equation}
\sigma_e \,:\,\sigma_{\mu}\,:\, \sigma_{\tau} \Leftrightarrow 1\,:\,0.16\,:\,0.16
\end{equation}
After scattering ends up, one should have an averaged statistical ensemble described by
\begin{equation}
1\,:\,0.16\,:\,0.16 \Leftrightarrow w_e \,:\,w_{\mu}\,:\,w_{\tau}
\label{234}
\end{equation}
where we have introduced the statistical weights $w_{\alpha}$, with $\alpha = $ $e$, $\mu$ and $\tau$
Since it does not correspond to the maximal statistical mixture, $w_e = w_{\mu} = w_{\tau}$, decoherence effects may lead the free streaming flavor ensemble to the maximal entropy configuration.

The composite quantum system of neutrino flavors certainly reaches the configuration of a maximal statistical mixture corresponding to $S = \ln(3)$ before entering the non-relativistic regime.
Due to some extrinsic decoherence mechanism, one identifies the variation of the von-Neumann entropy as a deviation from the entropy of such a maximal statistical mixing through $\delta S_{\mbox{\tiny VN}} = \ln(3) - S(\rho_{(t=0)})$, for which $t=0$ is defined as the time of neutrino decoupling and $S(\rho_{(t=0)})$ is computed in terms of the statistical weights $w_{\alpha}$.

\begin{figure}
\epsfig{file= 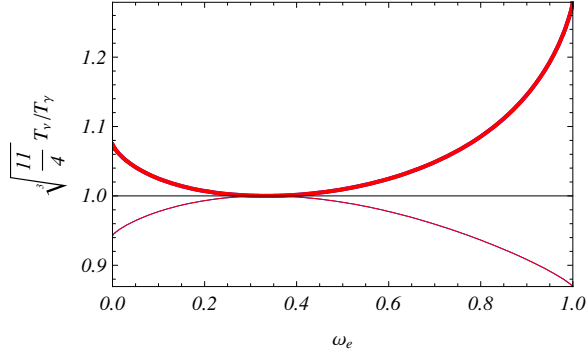, width= 8 cm}
%\vspace{-0.6 cm}
\caption{\label{an1} Upper (thick lines) and lower (thin line) maximal limits for modifications on the averaged ratio $T_{\nu\0}/T_{\gamma\0} = (4/11)^{1/3}$ due to maximal entropy changes $\delta S_{\mbox{\tiny VN}}$.
The flavor statistical mixing is parameterized by the electronic statistical weight, $w_e$.
We have considered a three-level composite quantum system with neutrino mixing angles approximated by the {\em tri-bimaximal}  parameters, i. e. $\theta_{12} = \arcsin{1/\sqrt{3}}$, $\theta_{23} = \pi/4$, and $\theta_{13} = 0$.
The unitarity condition, $w_{e} + w_{\mu} + w_{\tau} = 1$, with $w_{\mu}  = w_{\tau}$, reduces the dependence of the entropies on the statistical weights to the one-degree of freedom dependence on $w_{e}$. Blue and red lines describing the constraint of $\delta S_{\mbox{\tiny VN}}$ respectively with $S(\rho)$ and $S(\rho^{\prime})$ are coincident.}
%\vspace{-0.7 cm}
\end{figure}

The alternative way of quantifying such an entropy increasing after decoupling is through the parametrization of $\delta S_{\mbox{\tiny VN}}$ by $\delta S^{DL}_{\mbox{\tiny VN}} =  S(\langle \rho\rangle_{\rm time}) - S(\rho_{(t=0)})$
Assuming that some intrinsic (dynamical) decoherence mechanism suppresses the time dependence of the non-diagonal elements of the density matrix, the delocalization (DL) effects can be reproduced by time-averaging the density matrix: $\langle \rho \rangle_{\rm time}$,

Both parameterizations can be conceptually modified by adding to $\delta S^{DL}_{\mbox{\tiny VN}}$ the entropy change due to a {\em non-selective} measurement scheme given by $\Delta S (t = 0)$ from Eq.~(\ref{eq26A}).
As one can depict from Fig.~\ref{an1}, the {\em non-selective} measurements do not change $\delta S_{\mbox{\tiny VN}}$ at $t = 0$.
By assuming the next to standard phenomenological values for the neutrino flavor mixing angles, namely the {\em tri-bimaximal} approximation $\theta_{12} = \arcsin{(1/\sqrt{3})}$, $\theta_{23} = \pi/4$, and $\theta_{13} = 0$, one can depict from Fig.~\ref{an1} the modifications on the averaged ratio $T_{\nu\0}/T_{\gamma\0} = (4/11)^{1/3}$ due to entropy changes produced by flavor statistical mixing as functions of the electronic statistical weight, $w_e$.
Besides the unitarity given by Eq.~(\ref{eq05}), we have set $w_{\mu} = w_{\tau}$ in order to plot the curves.

Otherwise, by replacing $\delta S_{\mbox{\tiny VN}} =  \ln(3) - S(\rho_{(t=0)})$ by $\delta S^{DL}_{\mbox{\tiny VN}} =  S(\langle \rho\rangle_{\rm time}) - S(\rho_{(t=0)})$ into Eq.(\ref{xxx}), one quantifies the degree of mixture of the flavor definite statistical ensemble as function of its mixing properties since the  decoherence is exclusively due to delocalization effects, i. e. an intrinsic decoherence mechanism.

\begin{figure}
\epsfig{file= 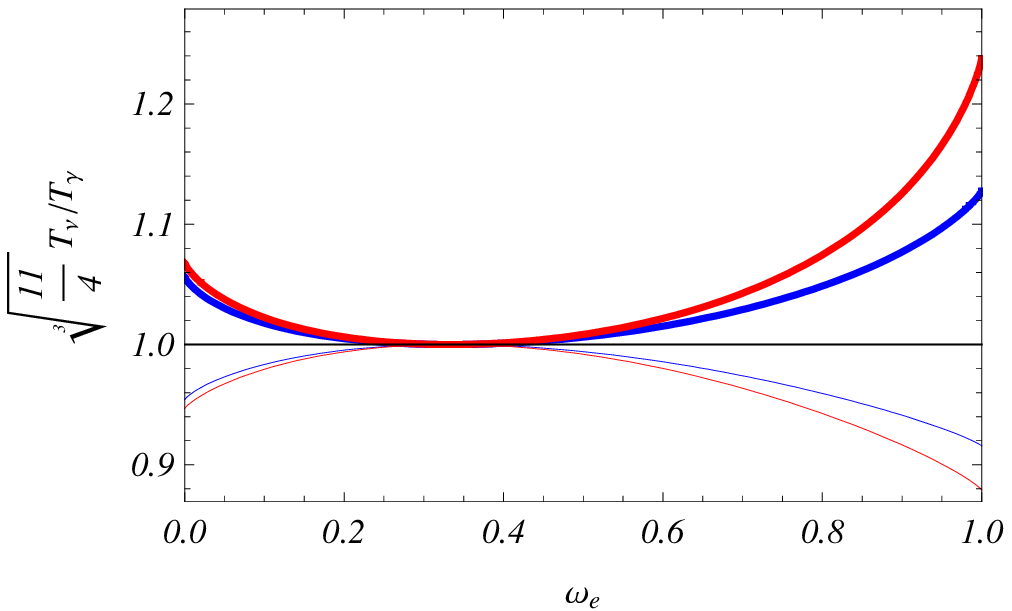, width= 8 cm}
\epsfig{file= 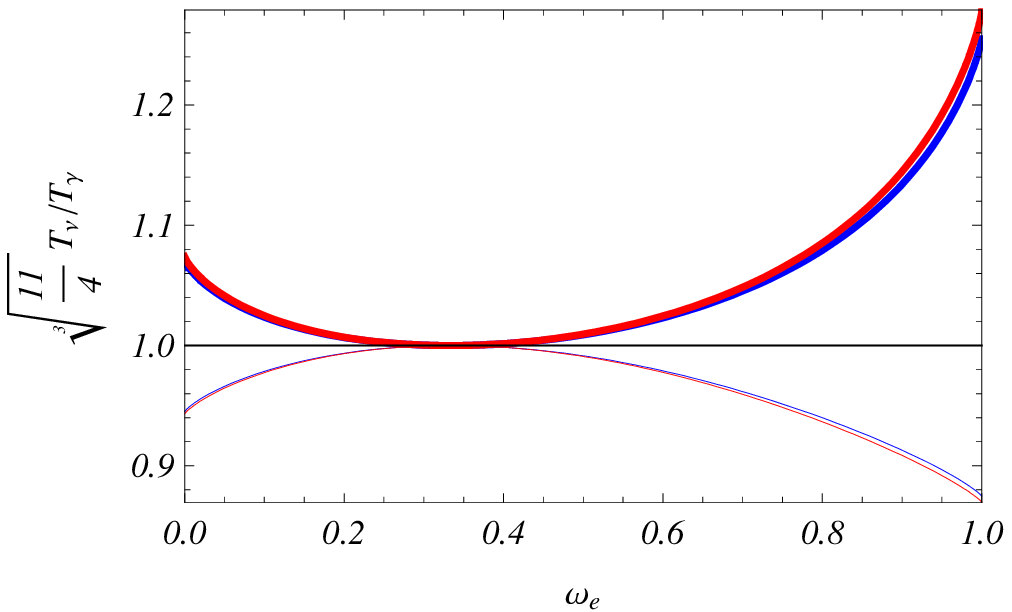, width= 8 cm}
\epsfig{file= 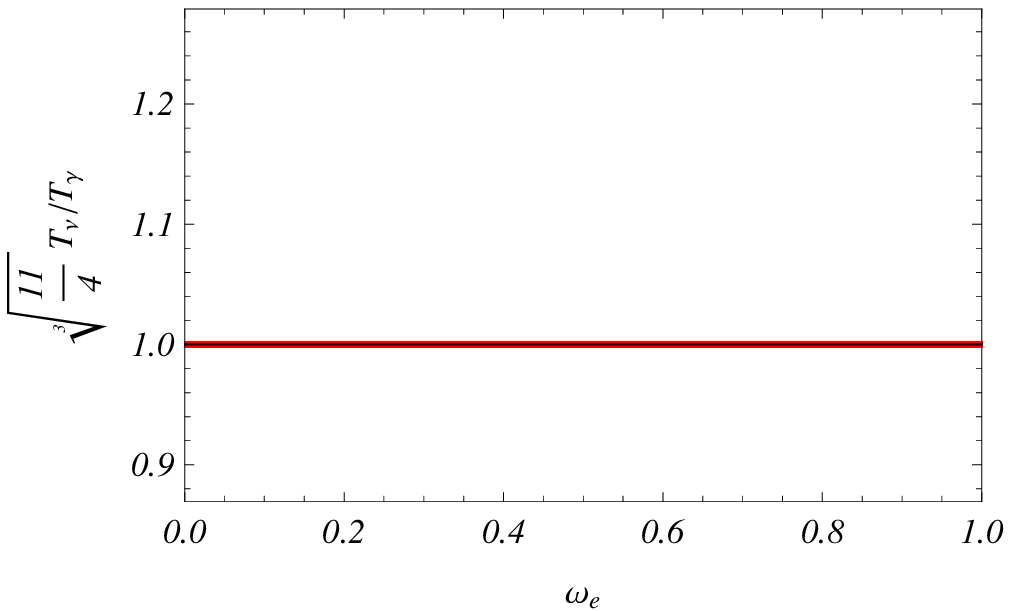, width= 8 cm}
\vspace{-0.6 cm}
\caption{\label{an2} Upper (thick lines) and lower (thin lines) limits for the modifications on the averaged rate $T_{\nu\0}/T_{\gamma\0} = (4/11)^{1/3}$ due to late-time entropy changes $\delta S^{DL}_{\mbox{\tiny VN}}$ produced
by intrinsic decoherence (delocalization) effects.
The mixing properties are defined in terms of the mixing angles.
In the first plot we have assumed the {\em tri-bimaximal} mixing with $\theta_{12} = \arcsin{1/\sqrt{3}}$, $\theta_{23} = \pi/4$, and $\theta_{13}= 0$.
In the second one we have set $\theta_{12} =\theta_{23} = \theta_{13}= \pi/4$.
And in the third one obviously no mixing is considered.
Red lines are obtained through the {\em non-selective} quantum measuremet scheme.}
%\vspace{-0.7 cm}
\end{figure}

The differences due to flavor mixing properties can be depicted from the Fig.~\ref{an2} for which we have considered three particular cases: i) the next to standard phenomenological values for the neutrino flavor mixing angles, namely the {\em tri-bimaximal} mixing approximation for which $\theta_{12} = \arcsin{1/\sqrt{3}}$, $\theta_{23} = \pi/4$, and $\theta_{13}  0$ ii) the maximal mixing for which $\theta_{12} =\theta_{23} = \theta_{13}= \pi/4$; and iii) no mixing for which $\theta_{12} =\theta_{23} = \theta_{13}= 0$.
In case of analyzing $\delta S_{\mbox{\tiny VN}}$ (c. f. Fig.~\ref{an2}) the entropy change ($\Delta S$) due to a {\em non-selective} measurement intervention is explicit.

To sum up the relevant aspects of the above properties,
Fig.~\ref{an1} show the upper and lower limits for the modifications on the rate $T_{\nu\0}/T_{\gamma\0} = (4/11)^{1/3}$ due to the corrections obtained from Eq.(\ref{xxx}) for the cases where $\delta S_{\mbox{\tiny VN}} = \ln(3) - S(\rho^{\prime})$ (red lines) and $\delta S_{\mbox{\tiny VN}} = \ln(3) - S(\rho)$ (blue lines).
Fig.~\ref{an2} follows the same correspondence.
As expected, Figs.~\ref{an1} and \ref{an2} reveals that the maximal decoherence effects occur for $w_e = 1$ which correspond to a pure state configuration at time of neutrino decoupling.
One also notices that quantum mixing is fundamental for introducing the additive quantum entropy.
It is convenient to notice that the entropy change due to a {\em non-selective} measurement performed over a maximal statistical mixture is null, i. e. $S(\rho) = S(\rho^{\prime})$.
For the values corresponding to the rapport from Eq.~(\ref{234}), one should have $w_{e}\sim 0.68$.
It leads to corrections of the order of $3$-$4\%$ on the maximal bounds.


%In addition to the above analysis, one could consider the variation of the entropy involved in the transition from {\em selective} to {\em non-selective} measurements that introduces the so-called mixing entropy,
%\begin{equation}
%\delta S_m =
%S\left(\sum_{\alpha} {P^{\alpha}_{(t)} \rho_{\alpha}}\right) - \sum_{\alpha} {P^{\alpha}_{(t)} S\left(\rho_{\alpha}\right)},
%\label{eq26C}
%\end{equation}
%that quantifies the difference between the entropy of a system projected by a {\em non-selective} quantum measurement, $S\left(\sum_{\alpha} {P^{\alpha}_{(t)} \rho_{\alpha}}\right)$, and the average of the entropies of the sub-ensembles, $\rho_{\alpha}$, described by states $M^{\alpha}_{(0)}$.
%When one sets $\rho_{\alpha} = M^{\alpha}_{(0)}$ for the {\em selective} measurement scheme, with $M^{\alpha}_{(0)}$ denoting the creation of a single-flavor state, the mixing entropy is reduced to
%\begin{equation}
%\delta S_m = S\left(\sum_{\alpha} {P^{\alpha}_{(t)} \rho_{\alpha}}\right),
%\label{eq26D}
%\end{equation}
%since $S\left(\rho_{\alpha}\right) = 0$.

%By redefining $\delta S^{DL}_{\mbox{\tiny VN}}$ as $\delta S^{DL}_{\mbox{\tiny VN}} - \delta S_m$, it should include the entropy changes due to {\em selective} measurement procedures which are not relevant at this point.
Finally, one should observe that the mixing entropy assumes its maximum value which, in case of an $n$-level system, corresponds to $\delta S_{m} = \ln\bb{n}$. For a three flavor system of neutrinos, it corresponds to $\delta S_{\mbox{\tiny VN}} = 0$.
For maximal statistical mixing, the {\em non-selective} measurement does not change neither the energy nor the entropy of the system while the {\em selective} measurement changes the entropy \cite{Breuer}.
It means that the von-Neumann entropy and the above-related quantities gain relevance in the study of the measurement procedures which take into account the flavor eigenstate correspondence to measurable energies.

%Turning back to the results from Eq.(\ref{xxx}), and
Observing that the entropy increasing follows the level of mixing of the systems, the maximal variation for the ratio $T_{\nu\0}/T_{\gamma\0} = (4/11)^{1/3}$ can be set through the reading of Eq.(\ref{xxx}) by means of Eq.~(\ref{sdsd}) that results in
\begin{equation}
\left|\left(1 - \frac{324}{7 \pi^{4}}\,\ln{3} \right)^{\frac{1}{3}} \right|
\lesssim
\frac{T_{\nu\0}}{T^{\prime}_{\nu\0}}
\lesssim
\left|\left(1 + \frac{324}{7 \pi^{4}}\,\ln{3} \right)^{\frac{1}{3}} \right|
~~\Rightarrow~~
0.86 \lesssim
\frac{T^{\prime}_{\nu\0}}{T_{\nu\0}}
\lesssim
1.28\label{Fim02}
\end{equation}
in case of realistic values given by the {\em tri-bimaximal} mixing.
The above values can be mitigated if one considers the $\delta S^{DL}_{\mbox{\tiny VN}}$ in place of $\delta S_{\mbox{\tiny VN}}$.

In the same fashion of some eventual indirect exotic coupling of neutrinos to electrons or photons that could have kept neutrinos longer in equilibrium with photons, entropy changes due to flavor mixing introduce a novel ingredient that suggests that the $\nu - \gamma$ number density ratio could not be diluted by $4/11$.
The possibility of attenuating the constraints on the late-time entropy production from the large scale structure and CMB anisotropies has already been considered \cite{33,34,35,Kohri}.
Herein the referred entropy modifications can change the rapport between the effective number of neutrino families, $N_{\nu}$, and any parameter phenomenologically depicted from the pattern of large scale structures and CMB anisotropies.

To clear up this point, let us assume that the standard value for the $\nu-\gamma$ energy density ratio, $7/8$, leads to the following relation between the red-shift of matter-radiation equality and $N_{\nu}$,
\begin{equation}
1 + z_{eq} \propto \left[1 + \frac{7}{8} \left(\frac{4}{11}\right)^{4/3}N_{\nu}\right]^{-1}.
\label{sds}
\end{equation}
To keep the ratio $\rho_{\nu}/\rho_{\gamma} = N_{\nu} (7/8) (4/11)^{4/3}$ consistent with the phenomenology, the modifications introduced by the growing von-Neumann entropy discussed above introduce the upper and lower bounds to $N_{\nu}$, through the modified parameter
\begin{equation}
N^{\prime}_{\nu} \sim N_{\nu} \left|\left(1 \pm \frac{324}{7 \pi^{4}}\,\delta S_{\mbox{\tiny VN}(w_e = 1)} \right)^{-\frac{4}{3}}\right|
\end{equation}
that, from this point, has to be interpreted as novel phenomenological parameter, and not as the number of neutrino species.
For $N_{\nu} \sim 3$, one should have the bounds
\begin{equation}
1.7 \lesssim N^{\prime}_{\nu} \lesssim 8.1,
\end{equation}
where the bounds are for neutrino ensembles being produced as pure states after decoupling and evolving to a maximal statistical mixture in the free-streaming regime.

Essentially, {\em we are not modifying the predictions for the number of neutrino species, $N_{\nu}$}, which was already accurately computed, for instance, at \cite{Dodd}, where the distortions in the $\nu_e$ and $\nu_{\mu}/\nu_{\tau}$ phase-space distribution that arise in the standard cosmology due to electron-positron annihilations have been considered.
Our results just relieve the constraints on the value of the parameter $N^{\prime}_{\nu}$ that re-enters into the expression derived from matter-radiation equality (c. f. Eq.~(\ref{sds})) and that could lead to some phenomenological tension \cite{02,144B}.

Our result enlarges the range of phenomenological agreement for tantalizing cosmological and terrestrial evidences that suggest the number of light neutrinos may be greater than three \cite{extra00,extra01}.
A recent re-examination of cosmological bounds on extra light species have been performed \cite{extra01} in order to consider the cosmological scenario with two sterile neutrinos and explore whether partial thermalization of the sterile states can mitigate the conflict between apparently ambiguous cosmological constraints on the number of neutrino species.

Accurately computed values of Helium abundance depicted from the Big Bang nucleosynthesis (BBN)formalism constrains the number of relativistic neutrino species present during nucleosynthesis, while measurements of the CMB angular power spectrum constrains the values of effective energy density of relativistic neutrinos and photons.
Therefore, scenarios where new sterile neutrino species may have different contributions to $N_{\nu}^{(eff)}$, respectively from BBN and CMB data, can be reconciled through the entropy corrections to C$\nu$B computed through the approach that we have introduced.
The same argument can be reported on the analysis of increasing the effective number of neutrino species, $N_{\nu}$, in the early universe, focussed on introducing extra relativistic species (hot dark matter or dark radiation \cite{extra02}).

The above results does not change the significance and the magnitude of the finite-temperature electromagnetic corrections to the energy density of the $\gamma e^+ e^-$ radiation plasma \cite{Dodd,142,143} or of the finite temperature QCD corrections \cite{Dolgov02}.
They are of the same order of magnitude of flavor mixing corrections upon the averaged temperature of decoupling for different neutrino species, which also depend on the mixing parameters \cite{Dolgov02}.
Although the above limits relieve the constraints on the possible values for $N_{\nu}$, neutrino heating/freezing corrections can be a little larger than those predicted by the finite temperature quantum field theories, for instance, late-time entropy production due to some weakly interacting scalar field decay \cite{01,02}.

To conclude, one should observe that it is commonly assumed that ultra-relativistic thermal relics are in perfect equilibrium state even after decoupling.
For photons in cosmic microwave background (CMB) this has been established with a very high degree of accuracy.
Thus, the same assumption has been made about free-streaming neutrinos.
Our line of reasoning calls into attention the premise of coherent flavor eigenstates at the epoch of neutrino decoupling from the radiation background.
It has been assumed that the free evolution of flavor ensembles leads to some spontaneous lowering of the coherence interference effects, associated with the destruction of the oscillation pattern and with the vanishing of the $3$-partite quantum entanglement.
Some previous studies have addressed to the issue of the decoherence history of the cosmological neutrinos and its implications on probing best values for neutrino masses and on modifying the power spectrum of large scale structures \cite{Fuller,Kis08,Dolgov02}.
In the single-particle quantum mechanics framework, it has been supposed that flavor wave-packets could have spatial extents that would be comparable to the space-time curvature scale of the universe itself.
The delocalization effects lead to decoherence in the same way as we have quoted in this letter (c. f. Fig.~\ref{an2}.)

Likewise, it could also result from some suitable prescription for some extrinsic dissipative mechanism that results into the wave function collapse, for instance, as a consequence of describing the flavor ensemble as an open quantum system.
Assuming that some kind of decoherence mechanism results in increasing the level of mixing for neutrino flavor ensembles in the cosmological background, the analysis developed here would not have only addressed to the entropy issues related to neutrinos, but would also suggest some new insights into the role of quantum coherence and decoherence in the history of these relic particles.

%It is relevant to notice that non-decaying particles created at thermal equilibrium are actually rather dense in phase space. %In addition, the intrinsic decoherence effect about which we have reported takes place very rapidly after the neutrino decoupling from the cosmological plasma.
%Extrinsic decoherence effects due to gravitational effects appears only at late times in the cosmological scale.
%Such aspects together make our approach sufficiently realistic for studying the mixing effects on the cosmological neutrino entropy.

Finally, our results states that the flavor quantum mixing of neutrino mass eigenstates associated to decoherence effects are fundamental for producing an additive contribution of quantum entropy to the cosmological neutrino thermal history.
According to our framework, it does not modify the predictions for the number of neutrino species, $N_{\nu} \approx 3$.
It can only relieve the constraints between $N_{\nu}$ and the neutrino to radiation temperature ratio, $T_{\nu}/T_{\gamma}$, by introducing a novel ingredient to re-direct the interpretation of some recent tantalizing evidence than $N_{\nu}$ is significantly larger than by more than $3$.
Obtaining the neutrino entropy changes well inside the free-streaming propagation regime is therefore a relevant aspect that has to be considered while computing cosmological neutrino properties, namely the cosmic energy density, the constraints related to the precise number of neutrino species, $N_{\nu}$, the specific entropy itself, and eventually, the neutrino mass values \cite{Bernardini}.

{\em Acknowledgments - This work has been supported by the Brazilian Agencies FAPESP (grant 12/03561-0) and CNPq (grant 300233/2010-8).}


\begin{thebibliography}{99}

\bibitem{01}
%Late-time entropy production due to the decay of domain walls
M. Kawasaki and F. Takahashi, Phys. Lett. {\bf B618}, 1 (2005).
\bibitem{02}
%Late-time Entropy Production from Scalar Decay and Relic Neutrino Temperature
P. Adhya, D. Rai Chaudhuri and S. Hannestad, Phys. Rev. {\bf D68}, 083519 (2003).
\bibitem{03}
%Cosmological Constraints on Late-time Entropy Production
M. Kawasaki, K. Kohri and N. Sugiyama, Phys. Rev. Lett. {\bf 82}, 4168 (1999).
\bibitem{04}
%Late-Time Entropy Production and Relic Abundances of Neutralinos
T. Nagano, M. Yamaguchi, Phys. Lett. {\bf B438}, 267 (1998).
\bibitem{144}
A. F. Heckler, Phys. Rev. {\bf D49}, 611 (1994).
\bibitem{Bernardini}
A. E. Bernardini and V. A. S. V. Bittencourt, Astroparticle Phys. {\bf 41}, 31 (2013).
\bibitem{Dodelson}
S. Dodelson, {\em Modern Cosmology: Anisotropies and Inhomogeneities in the Universe}, (Academic Press, New York, 2003).
\bibitem{Ma94}
C. P. Ma and E. Bertschinger, Astrophys. J. {\bf 455}, 7 (1995).
\bibitem{Breuer}
H. P. Breuer and F. Petruccione, {\em The Theory of Open Quantum Systems} (Oxford University Press, New York, 2002).
\bibitem{33}
A. G. Riess {\it et al.}, Astron. J. {\bf 116}, 1009 (1998).
\bibitem{34}
S. Perlmutter {\it et al.}, Astrophys. J. {\bf 517}, 565 (1999).
\bibitem{35}
S. Hancock, G. Rocha, A. N. Lasenby and C. M. Guti\'{e}rrez, Mon. Not. R. Astron. Soc. {\bf 294}, L1 (1998);
G. Efstathiou {\it et al.}, Mon. Not. R. Astron. Soc. {\bf 393}, L47 (1999);
M. Tegmark, Astrophys. J. {\bf 514}, L69 (1999).
\bibitem{Kohri}
M. Kawasaki, K. Kohri and N. Sugiyama, Phys. Rev. {\bf D62}, 023506 (2000).
\bibitem{Fuller}
G. M. Fuller and C. T. Kishimoto, Phys. Rev. Lett.{\bf 102}, 201303 (2009).
\bibitem{Kis08}
C. T. Kishimoto and G. M. Fuller, Phys. Rev. {\bf D78}, 023524 (2008).
\bibitem{Dolgov02}
A. D. Dolgov, Phys. Rept. {\bf 370}, 333 (2002).

%\bibitem{Boy10}
%A. Boyarsky, A. Neronov, O. Ruchayskiy, and I. Tkachev, Phys. Rev. Lett. {\bf 104}, 191301 (2010).
%\bibitem{Pas06}
%J. Lesgourgues and S. Pastor, Phys. Rept. {\bf 429}, 307 (2006).
%\bibitem{10}
%V. B. Braginsky and F. Ya. Khalili, {\em Quantum Measurement} (Cambridge University Press, Cambridge, 1992).
%\bibitem{20}
%E. B. Davies, {\em Quantum Theory of Open Systems} (Academic Press, London, 1976).
%\bibitem{30}
%E. B. Kraus, {\em Quantum Theory of Open Systems} (Academic Press, London, 1983).
\bibitem{Dodd}
S. Dodelson and M. S. Turner, Phys.Rev. {\bf D46}, 3372 (1992).
\bibitem{142}
R. Maartens and J. Triginer, Gen. Rel. Grav. {\bf 32}, 1711 (2000).
\bibitem{143}
B. D. Fields, S. Dodelson and M. S. Turner, Phys. Rev. {\bf D47}, 4309 (1993).
\bibitem{144B}
S. Hannestad,  Prog. Part. Nucl. Phys. {\bf 65}, 185 (2010).
\bibitem{extra00} 
Z. Hou {\em it. et al.}, {\em Constraints on Cosmology from the Cosmic Microwave Background Power Spectrum of the 2500-square degree SPT-SZ Survey}, arXiv:1212.6267 [astro-ph].
\bibitem{extra01} 
T. D. Jacques, L. M. Krauss and C. Lunardini, {\em Additional Light Sterile Neutrinos and Cosmology}, arXiv:1301.3119 [astro-ph].    
\bibitem{extra02} 
C. Boehm, M. J. Dolan and C. McCabe, {\em Increasing Neff with particles in thermal equilibrium with neutrinos}, arXiv:1207.0497 [astro-ph].
\end{thebibliography}
    \end{document}